\documentclass[twocolumn,superscriptaddress,floatfix,preprintnumbers,prl ,hyperref]{revtex4-2}
\usepackage[colorlinks=true,breaklinks=true]{hyperref}
\usepackage{comment}
\usepackage[utf8]{inputenc}
\hypersetup{allcolors=[rgb]{0.0 0.0 0.6},linkcolor=[rgb]{0.75 0.05 0.05}}
\usepackage{mathtools}
\usepackage[dvipsnames]{xcolor}
\usepackage{float}
\usepackage{letltxmacro}
\LetLtxMacro{\oldcite}{\cite}
\renewcommand{\cite}[1]{\mbox{\oldcite{#1}}}

\usepackage{orcidlink}

\long\def\exclude#1{}

\newcommand{\vs}{v_{\rm s}}
\newcommand{\rs}{r_{\rm s}}
\newcommand{\rh}{r_{\rm h}}
\newcommand{\rd}{r_{\rm d}}
\newcommand{\dSN}{d_{\rm SN}}
\newcommand{\gd}{\gamma_{\rm d}}
\newcommand{\Td}{T_{\nu,{\rm d}}}

\bibpunct{[}{]}{,}{n}{}{,}

\begin{document}

\title{Large Neutrino Secret Interactions, Small Impact on Supernovae}

\author{Damiano F.\ G.\ Fiorillo \orcidlink{0000-0003-4927-9850}} 
%\email{damianofg@gmail.com}
\affiliation{Niels Bohr International Academy, Niels Bohr Institute,
University of Copenhagen, 2100 Copenhagen, Denmark}

\author{Georg G.\ Raffelt
\orcidlink{0000-0002-0199-9560}}%\email{raffelt@mpp.mpg.de}
\affiliation{Max-Planck-Institut f\"ur Physik (Werner-Heisenberg-Institut), F\"ohringer Ring 6, 80805 M\"unchen, Germany}

\author{Edoardo Vitagliano
\orcidlink{0000-0001-7847-1281}}%\email{edoardo@physics.ucla.edu}
\affiliation{Racah Institute of Physics, Hebrew University of Jerusalem, Jerusalem 91904, Israel}

\date{July 31, 2023, revised October 24, 2023}

\begin{abstract}
When hypothetical neutrino secret interactions ($\nu$SI) are large, they form a fluid in a supernova (SN) core, flow out with sonic speed, and stream away as a fireball. For the first time, we tackle the complete dynamical problem and solve all steps, systematically using relativistic hydrodynamics. The impact on SN physics and the neutrino signal is remarkably small. For complete thermalization within the fireball, the observable spectrum changes in a way that is independent of the coupling strength. One potentially large effect beyond our study is quick deleptonization if $\nu$SI violate lepton number. By present evidence, however, SN physics leaves open a large region in parameter space, where laboratory searches and future high-energy neutrino telescopes will probe $\nu$SI.
\end{abstract}

\maketitle

{\bf\textit{Introduction.}}---The cosmic dark-matter problem, the baryon asymmetry of the universe, the CP problem of QCD, and the unknown origin and nature of neutrino masses all suggest physics beyond the particle-physics Standard Model. One portal to new particle physics may be provided by hitherto unknown interactions among neutrinos~\cite{Berryman:2022hds}, with effects that are notoriously difficult to measure. Such neutrino secret interactions ($\nu$SI) must be mediated by a new force carrier of unknown spin parity and mass, and could conserve or violate lepton number. It has long been held \cite{Dicus:1982dk, Gelmini:1982rr, Kolb:1987qy, Manohar:1987ec, Berezhiani:1987gf, Dicus:1988jh, Fuller:1988ega, Berkov:1988sd,Konoplich:1988mj, Berezhiani:1989za, Farzan:2002wx, Heurtier:2016otg, Blennow:2008er, Das:2017iuj, Shalgar:2019rqe, Chang:2022aas, Fiorillo:2022cdq,Akita:2022etk, Cerdeno:2023kqo} that a natural test bed should be core-collapse supernova (SN) physics that is famously dominated by neutrinos \cite{Janka:2012wk, Janka2017Handbooka, Burrows+2020, Vitagliano:2019yzm, Mezzacappa+2020, Burrows+2021}.

If coupling $g_\phi$ and mass $m_\phi$ of the new force carrier $\phi$ are small enough, the main effect is energy loss by $\phi$ radiation, providing the traditional cooling bounds based on the SN~1987A neutrino signal~\cite{Farzan:2002wx, Heurtier:2016otg}. For larger masses, $\phi\to\nu\nu$ decays would provide many 100-MeV-range neutrinos, representative of the SN core, in conflict with SN~1987A data~\cite{Fiorillo:2022cdq}, and may be investigated also with future galactic SNe \cite{Akita:2022etk}. However, the exclusion region in the $g_\phi$--$m_\phi$ plane has a ceiling at $g_\phi$ so large 
($g_\phi m_\phi\gtrsim 10^{-7}\,\rm MeV$ for $m_\phi\gtrsim 1\,\rm MeV$) that the $\phi$ would be trapped inside the protoneutron star (PNS). At even larger couplings, such that the neutrino-neutrino mean free path (MFP) is shorter than the PNS radius, several new phenomena emerge.

One possibly dramatic effect arises when $\nu$SI violate lepton number as in the traditional majoron models. As much as 0.30 leptons per baryon are initially trapped, providing a large electron chemical potential, and causing the SN core to be rather cold after collapse. Lepton number usually escapes by diffusion and convection over a few seconds. On the other hand, quick deleptonization by $\nu\nu\to\bar{\nu}\bar{\nu}$, partly already during infall, modifies the entire SN paradigm. Still, in contrast to what is sometimes stated, such a scenario is not necessarily excluded because the hydrodynamic shock wave could arise from a thermal bounce \cite{Fuller:1988ega, Rampp:2002kn}. These are riveting questions that need addressing in selfconsistent SN~simulations. 

In this {\em Letter} we focus on effects unrelated to lepton-number violation: the reputedly large $\nu$SI impact on neutrino transport. In an early paper, Manohar suggested that $\nu$SI would make neutrinos diffuse in a gas of each other, retarding their flow, and thus violate the SN~1987A burst duration \cite{Manohar:1987ec}. This misconception was countered by Dicus et al.\ \cite{Dicus:1988jh} who stressed that strongly coupled neutrinos form a relativistic fluid and studied free expansion after sudden release. Recently, Chang et al.\ \cite{Chang:2022aas} have revived this long-dormant topic and advanced two scenarios of neutrino-fluid evolution, dubbed ``burst outflow'' and ``wind outflow,'' corresponding respectively to sudden release and steady emission. They questioned if special conditions were needed to realize the latter and if it could occur at all. Their main message was that burst outflow would lead to large observable effects mainly by extending the SN burst duration.

To develop an unambiguous answer, we immediately dismiss burst outflow because the sudden release of a fluid ball bears no resemblance to quasi-thermal emission by the protoneutron star (PNS) over several seconds. It creates a ball $10^5$ times the PNS radius of some 10~km. Moreover, it has long been known \cite{vitello1976hydrodynamic, yokosawa1980relativistic} that a suddenly released ball of relativistic fluid, after a short transient, behaves like a fireball: a constant-thickness shell that expands with the speed of light. The neutrino burst would not lengthen. (We elaborate on the sudden-release fireball solution in the Supplemental Material~\cite{supplementalmaterial}.) On the other hand, while steady wind looks plausible, it sidesteps the question of how it would dynamically arise after SN collapse, and it cannot be related to any physical observable at Earth.

\begin{figure*}
    \centering
    \includegraphics[width=\textwidth]{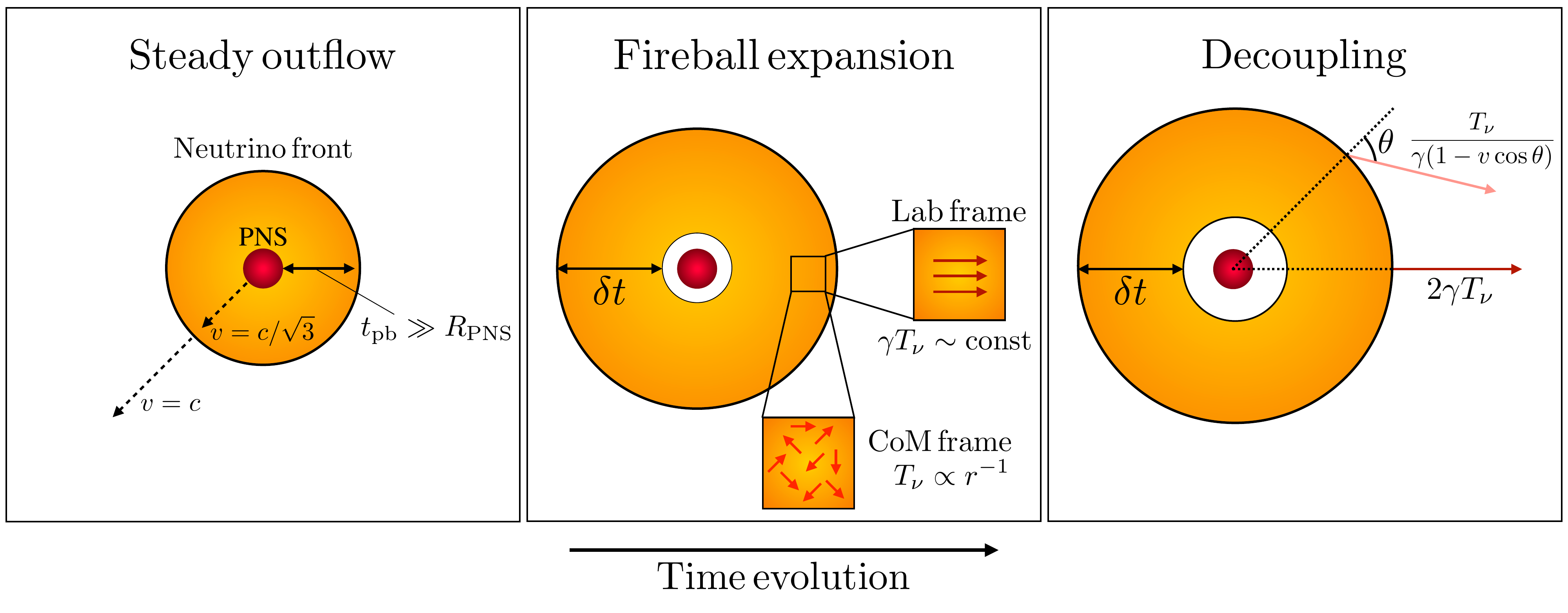}
%     \vskip-6pt
    \caption{Schematic evolution of neutrino fluid emission. \textit{Left panel:} Quasi steady emission from PNS with speed of sound $\vs=c/\sqrt{3}$, front of emission has reached distance $c\,t_{\rm pb}$ (post bounce time $t_{\rm pb}$). \textit{Middle panel:} The shell of neutrinos streams as a fireball, maintaining constant thickness $c\delta t$ (burst duration $\delta t$). In their comoving frame, neutrinos constantly isotropize by the $\nu$SI; in the lab frame, they are boosted and move collinear within an angle $\gamma^{-1}$. \textit{Right panel:} After they decouple, neutrinos maintain their spectrum. Neutrinos moving radially towards the observer have a large effective temperature of order $2\gamma T_\nu$, while neutrinos from the edges have a much lower temperature due to the Lorentz boosting.}
    \label{fig:sketch}
%    \vskip-6pt
\end{figure*}

For the first time, we tackle the full dynamical problem in spherical geometry with physical boundary conditions. A simplified source model with a beginning and end of thermal emission spawns a \textit{dynamical} solution with a luminal front expanding into space. Locally near the PNS, it relaxes to steady emission similar wind outflow, and finally morphs to a fireball with constant thickness (see Fig.~\ref{fig:sketch} for a sketch). 

The idea circulates that several observables---neutrino average energy \cite{Shalgar:2019rqe}, time of arrival \cite{Manohar:1987ec}, and signal duration \cite{Chang:2022aas}---depend on $R_{\rm PNS}/\lambda_{\nu\nu}$, where $R_{\rm PNS}$ is the proto-neutron star radius and $\lambda_{\nu\nu}$ the $\nu\nu$ MFP. In contrast, we find no strong such dependence. The signal duration and flux spectrum are astonishingly similar to the standard case, although tens-of-percent effects may persist and influence both SN physics and high-statistics observations. The groundwork for our study is laid out in a detailed theoretical companion paper~\cite{Fiorillo:2023cas}. We always use natural units with $c=\hbar=k_{\rm B}=1$.

{\bf\textit{Setup of the problem.}}---As $\lambda_{\nu\nu}\ll R_{\rm PNS}$, the key premise is treating neutrinos as a relativistic fluid, where $p=\rho/3$ with $p$ and $\rho$ the comoving pressure and energy density. The stress-energy tensor is $T^{\mu\nu}=\frac{4}{3}\rho\,u^\mu u^\nu-\frac{1}{3}\rho g^{\mu\nu}$ with $u^\mu$ the bulk velocity. The general hydrodynamical equations are \cite{Weinberg:1972kfs} $\partial_\nu T^{\mu \nu}=S^\mu$, including a source term on the right-hand side for the exchange of energy and momentum with the background medium. In free space, $S^\mu=0$, and then these equations simply express the local conservation of energy and momentum. In spherical symmetry, there is only the radial velocity $v$ such that $u^0=\gamma=(1-v^2)^{-1/2}$ and $u^r=\gamma v$. There remain only two hydrodynamical equations. One is for the
lab-frame energy density $e(r,t)$
\begin{equation}\label{eq:sphere_time_equation-a}
   \partial_t e+\frac{\partial_r(e\xi r^2)}{r^2}
=\frac{e_{\rm eq}-e}{\lambda_{\nu N}}, 
\end{equation}
where $e_{\rm eq}$ is the energy density when the fluid is in local thermal equilibrium (LTE) with the nuclear medium, $\lambda_{\nu N}$ the neutrino MFP for absorption, and $\xi={4 v}/{(3+v^2)}$ a modified velocity variable that gives us the energy flux when it multiplies the energy density. A second equation is for momentum
\begin{eqnarray}
\kern-2em\partial_t(e\xi)&+&\frac{1}{3r^2}\partial_r\left[e\left(5-2\sqrt{4-3\xi^2}\right) r^2\right]
\nonumber\\[1.5ex]
\label{eq:sphere_time_equation-b}
&&\kern3.5em{}-\frac{2e}{3r}\left(\sqrt{4-3\xi^2}-1\right)
=-\frac{e\xi}{\lambda_{\nu N}}.
\end{eqnarray}
In contrast to a kinetic treatment, neutrinos as particles do not appear in the complete set of 
Eqs.~\eqref{eq:sphere_time_equation-a} and \eqref{eq:sphere_time_equation-b} for the functions $e(r,t)$ and $\xi(r,t)$ or $v(r,t)$.

In the comoving frame, neutrinos are isotropic with the energy density $\rho=3 e/(4\gamma^2-1)$. The distribution is thermal with separate chemical potentials for $\nu$ and $\bar\nu$ if number-changing processes $\nu\overline{\nu}\to\nu\overline{\nu}\nu\overline{\nu}$ are slow, or chemical equilibrium with $\mu_\nu=-\mu_{\bar\nu}$ if they are fast, or $\mu_\nu=\mu_{\bar\nu}=0$ if collisions violate lepton number. If the fluid cannot internally establish chemical equilibrium, lab-frame number densities $N(r,t)$ are conserved other than by exchange with the background according to  
\begin{equation}\label{eq:sphere_time_equation-c}
\partial_t N+\frac{\partial_r(N v r^2)}{r^2}
=\frac{N_\mathrm{eq}-N}{\lambda_{\nu N}}.
\end{equation}
$N_{\rm eq}$ obtains in local chemical and thermal equilibrium with the background. To solve a physical problem, these equations must be complemented with appropriate initial and/or boundary conditions. 

{\bf\textit{Energy transport in the PNS.}}---Before energy can be radiated into space, it must be transported to the PNS surface. The usual diffusion flux is $F=-(\bar\lambda/3)\nabla e_\mathrm{eq}$, where $\bar\lambda$ is the Rosseland average neutrino MFP. How is this affected by $\nu$SI? With or without them, the right-hand side of Eq.~\eqref{eq:sphere_time_equation-b} is a force balanced by neutrino pressure, and $F$ turns out to be the same \cite{Fiorillo:2023cas}. This conclusion was also mentioned in Ref.~\cite{Cerdeno:2023kqo} based on momentum conservation in neutrino collisions; however, since the energy flux comes out of balance with neutrino-nucleon collisions, which do not conserve the neutrino momentum, our conclusion can only be reached using the hydrodynamical approach. 

Between collisions with the medium, neutrinos thermalize in the fluid frame, so $\bar\lambda$ denotes a somewhat different average if $\lambda_{\nu N}$ depends on energy \cite{Fiorillo:2023cas}; for quadratic energy dependence it is about 2/3 smaller. Moreover, $e_\mathrm{eq}$ includes all flavors, so $\bar\lambda$ is a flavor average, and if $m_\phi$ is small enough, $\phi$ also contributes to $e_\mathrm{eq}$. Therefore, $\nu$SI change the exact mean opacity, but on the other hand, PNS cooling strongly depends on convection \cite{Epstein1979, Burrows+1988, Keil+1996, Janka+2001proc, Dessart+2006, Nagakura+2020}. Therefore, we worry less about how energy streams up from deeper layers and focus on fluid decoupling near the surface.

\begin{figure}
    \centering
    \includegraphics[width=0.9 \columnwidth]{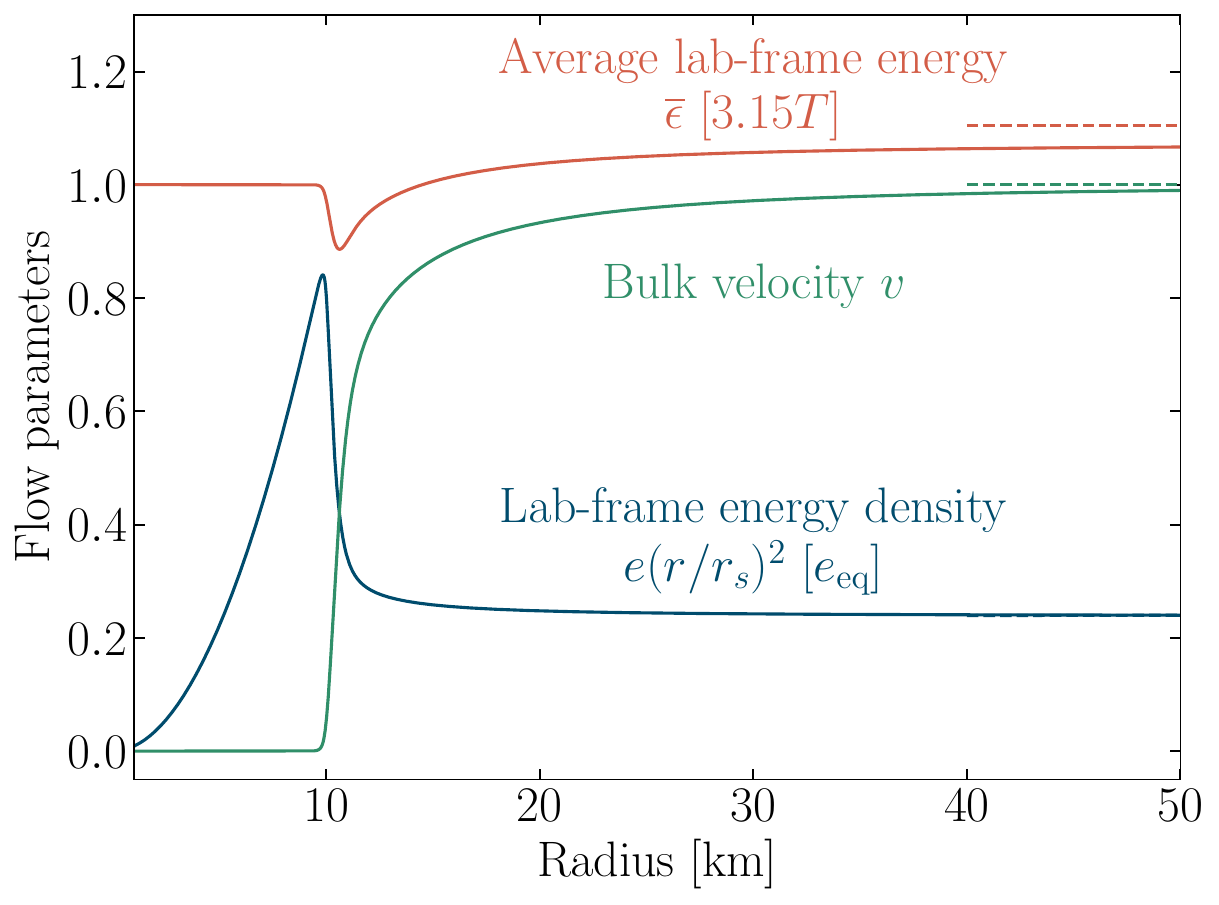}
     \vskip-6pt
    \caption{Flow parameters for steady fluid emission from an isothermal blackbody sphere (temperature $T$), using a smoothed surface at $\rs=10~$~km as described in the text.    Asymptotic values for $r\to\infty$ as dashed lines. The average lab-frame neutrino energy $\bar\epsilon$ assumes full efficiency for number-changing $\nu$SI interactions.}
    \label{fig:flow_parameters}
   \vskip-6pt
\end{figure}

{\bf\textit{Steady emission.}}---To get a first sense, we begin with a stationary solution for a simplified emission model: an isothermal sphere (radius $r_{\rm s}$, temperature~$T$) and energy-independent $\lambda_{\nu N}$. In Fig.~\ref{fig:flow_parameters} we show the resulting flow parameters for a numerical solution with $r_{\rm s}=10$~km and $\lambda_{\nu N}=0.2$~km for $r<r_{\rm s}$, and increasing with  $e^{6\,(r-\rs)/\mathrm{km}}$ for larger $r$ mainly to avoid a step function. $T$ is represented by $e_\mathrm{eq}$ constant inside, and outside following the same suppression profile. The fluid accelerates near the surface and quickly reaches luminal speed, similar to free-streaming neutrinos that become ever more collinear after a distance of a few $r_{\rm s}$. In the fluid, neutrinos are locally isotropic and thermal in the comoving frame with an ever decreasing $T_\nu$, whereas in the lab frame, most of them stream away radially.

A steady-state solution is unphysical because there must have been a beginning of emission and concomitant luminal wave front at a large distance. In our companion paper \cite{Fiorillo:2023cas}, we circumvent this issue with an outer shell at vanishing $T$ that absorbs the radiation. Between the shells, the analytical velocity profile is given by
\begin{equation}
    v(1-v^2)=\frac{2}{3\sqrt{3}}\left(\frac{\rs}{r}\right)^2,
\end{equation}
assuming a hard surface at $\rs$. At the surface, the fluid emerges with the speed of sound
$\vs=1/\sqrt{3}=0.577$, a general result also found in Ref.~\cite{Chang:2022aas}.
 With increasing radius, $v$ rapidly rises from $\vs$ to $c$ and asymptotically reaches a Lorentz factor
\begin{equation}\label{eq:gamma_factor}
    \gamma\simeq\frac{3^{3/4}}{\sqrt{2}}\,\frac{r}{\rs}
    \simeq 1.61\,\frac{r}{\rs}.
\end{equation}
Analogously, the comoving energy and number densities reach asymptotic values of
$\rho\simeq(0.21\,e_\mathrm{eq}/3)(\rs/r)^4$ and $n\simeq (0.29/3^{3/4})\,n^\mathrm{th}(\rs/r)^3$, the lab-frame energy density $e\simeq 4\gamma^2\rho/3\propto r^{-2}$.

From the numerical solution (with a slightly softened surface), we see in Fig.~\ref{fig:flow_parameters} how $e(r)$ quickly drops from within the source body to its asymptotic behavior. Together with increasing $v$, the energy flux $F=\xi e$ is conserved outside the source. Surprisingly, the blackbody emits an energy flux that is numerically only 3--4\% smaller than $F_{\rm bb}=e_{\rm th}/4$ given by the Stefan-Boltzmann law for standard neutrino radiation. We have no fundamental explanation why blackbody emission of a fluid should be so similar, yet not identical, to that of a gas. What ever the answer to this conceptual question, it defines the practical boundary condition at the source.

Our stationary solution looks similar to wind outflow of Ref.~\cite{Chang:2022aas}, with the crucial difference that our fluid is steadily produced by the source, not fed by a reservoir of trapped neutrinos that would be quickly exhausted. Our treatment of the boundary connects the thermal properties of the PNS to those of the escaping fluid.

In the comoving frame, the fluid is in equilibrium and thus characterized by its internal $T_\nu$.
If the source emits $\nu$ and $\bar\nu$ equally, and if number-changing processes by $\nu$SI are fast so neutrinos reach internal chemical equilibrium, the asymptotic $\rho$ implies $T_\nu=0.514\,T\,(\rs/r)$, a drop that is compensated by the work needed for the expansion, or equivalently, the increasing bulk radial motion: the particles simply become more collinear. The increasing $\gamma$ implies a constant asymptotic lab-frame neutrino \hbox{energy} of $\overline{\epsilon}=3.48\,T$. Without $\nu$SI, the lab-frame spectrum is thermal with the emitter's $T$ and thus  $\overline{\epsilon}=3.15\,T$.

If number-changing reactions are not in equilibrium, but still equal $\nu$ and $\bar\nu$ emission by the source, a nonvanishing degeneracy parameter $\eta=\mu_\nu/T_\nu$ develops, based on number and energy flux conservation. The asymptotic values are $\eta=-0.363$, $T_\nu=0.561\,T\, (\rs/r)$, and lab-frame $\overline{\epsilon}=3.75\,T$.  Constant $\eta$ implies entropy conservation (adiabatic expansion) once the fluid has settled into steady motion at $r$ larger than a few~$\rs$.

{\bf\textit{Relaxation to steady state.}}---If thermal emission begins suddenly at SN collapse, a luminal neutrino fluid front is launched into space. While physically plausible, we have explicitly checked numerically that this dynamical solution asymptotically approaches steady outflow near the source \cite{Fiorillo:2023cas}. Though our treatment of the source is schematic, we are confident that it correctly captures the transient. The main innovation is to include energy exchange with the nuclear medium, allowing the PNS to act as an energy reservoir that feeds neutrino emission for a long time compared with the PNS size, completely different from sudden fluid release, a concept that would only apply to the initial front wave.

{\bf\textit{Neutrino fireball expansion.}}---As the PNS cools, neutrino emission drops. After $\delta t$ of a few seconds, a shell of width $\delta t$ has been emitted. The subsequent evolution is well understood in the context of fireballs in gamma-ray bursts~\cite{Piran:1993jm} (see also Refs.~\cite{vitello1976hydrodynamic, yokosawa1980relativistic} for early theoretical studies and Refs.~\cite{Diamond:2023scc, Diamond:2023cto} for particle bounds from astrophysical transients). Since most of the fluid moves with $v\simeq 1$, the shell thickness cannot change, 
but its radius gradually expands. Within the shell, our steady-state flow parameters remain valid. 

This is also seen because, in steady state, a disturbance in the fluid travels with $\vs$ in the comoving frame, the latter however accelerating with increasing radius. There is a sound horizon $\rh\simeq 1.13\,\rs$ \cite{Fiorillo:2023cas}. The fluid at larger $r$ is unaware of anything happening at the emission surface, such as the source turning off. 

As the thickness does not change, the fireball is not a self-similar solution because it contains a characteristic length $\delta t$.
In fact, for free expansion of a relativistic gas, self-similar solutions with regular behaviors do not seem to exist. In our case, regular behavior is attained because, for energy injection over a period $\delta t$, the system always keeps memory of the scale $\delta t$ (see, e.g., Refs.~\cite{vitello1976hydrodynamic, yokosawa1980relativistic}), even after a time $t\gg \delta t$.

{\bf\textit{Observable neutrino signal.}}---Within the fireball, neutrinos possess a boosted blackbody spectrum. However, at some radius $\rd$, the density is so low that 
$\nu$SI decouple and then neutrinos stream freely. The large Lorentz factor of Eq.~\eqref{eq:gamma_factor} reveals that we observe neutrinos with nearly the same angular spread for both free streaming or fluid propagation, so time-of-flight effects are minimal. We thus picture the observable flux at a distance $\dSN\gg\rd$ to be steadily emitted by a spherical shell of radius $\rd$, taken to be 
the same for all energies.

Thus at a large distance one observes the superposition of the boosted blackbody spectra from each point on that sphere (right panel in Fig.~\ref{fig:sketch}). While the comoving $\Td$ and Lorentz factor $\gd$ are the same for all points, they are seen under different angles, producing different spectra due to Doppler boosting. The limb of the sphere looks much colder (effective temperature of order $\Td/\gd$) than the center (effective temperature of order $2 \Td \gd$).
Explicitly, the superposed number flux spectrum for a single species is found to be~\cite{Diamond:2023scc}
\begin{equation}
\frac{d \Phi}{d\epsilon}=\frac{\rd^2}{4\pi^2\dSN^2} \frac{\tilde{T}\epsilon}{\gd^2}\log\left[1+e^{\eta-\epsilon/2\tilde{T}}\right],
\end{equation}
where $\tilde{T}=\gamma^d T_\nu^d=\gamma T_\nu$ if we recall the constancy of $\gamma T_\nu$ during fireball expansion. When $\nu$SI number-changing reactions are in equilibrium near the PNS, $\tilde{T}=0.828\,T$ and $\eta=0$. For the opposite case of number-conserving dynamics, $\tilde{T}=0.903\,T$ and $\eta=-0.363$. 

Finally, using the Lorentz factor of Eq.~\eqref{eq:gamma_factor}, the observer spectrum is 
\begin{equation}\label{eq:fluidspectrum}
    \frac{d\Phi}{d\epsilon}=\frac{\rs^2}{6\sqrt{3}\pi^2 \dSN^2}\tilde{T}\epsilon\log\left[1+e^{\eta-\epsilon/2\tilde{T}}\right].
\end{equation}
Nothing depends on $\rd$, justifying our earlier assumption of taking decoupling to be instantaneous. This is to be compared with the blackbody radiation without $\nu$SI
\begin{equation}\label{eq:bbspectrum}
    \left(\frac{d\Phi}{d\epsilon}\right)_{\mathrm{bb}}=\frac{\rs^2}{8\pi^2 \dSN^2}\frac{\epsilon^2}{e^{\epsilon/T}+1}.
\end{equation}
One can check that the integrated energy flux from Eq.~\eqref{eq:fluidspectrum} is indeed 0.96 the one from Eq.~\eqref{eq:bbspectrum}, thus maintaining exactly the energy outflow stated earlier~\cite{Fiorillo:2023cas}.

\begin{figure}[b]
    \centering
    \includegraphics[width=0.9\columnwidth]{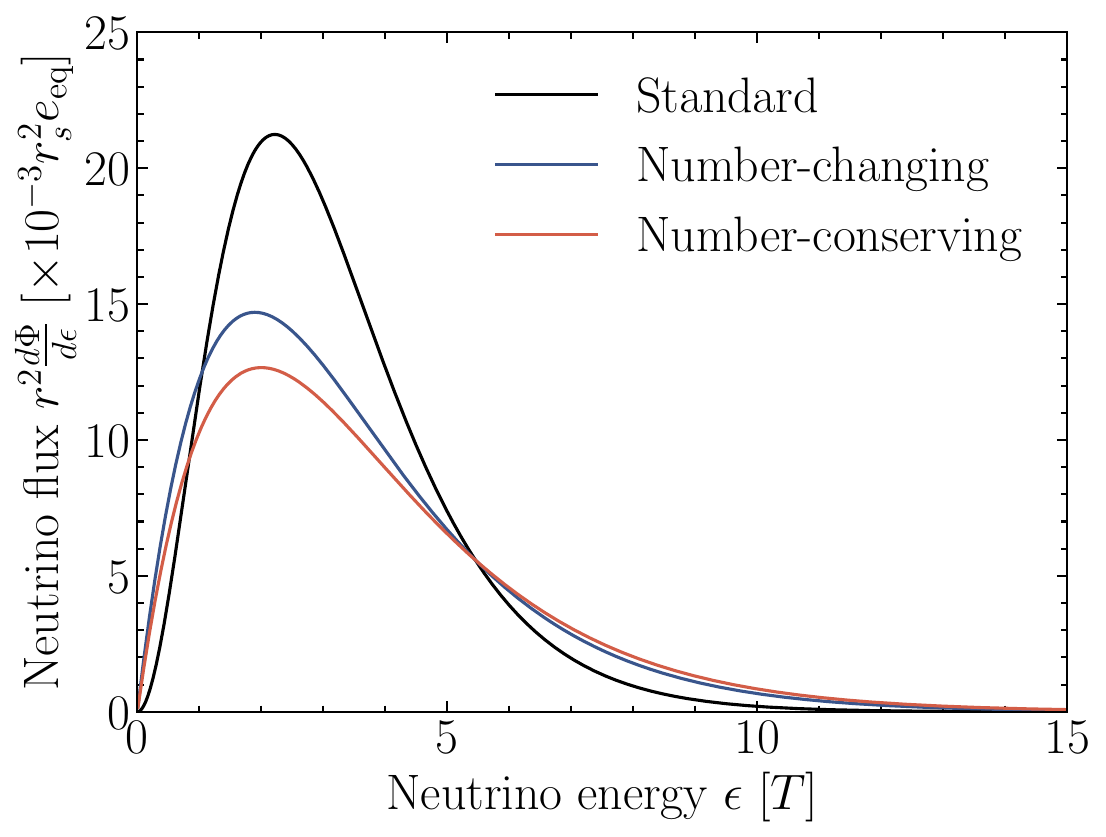}
%    \vskip-6pt
    \caption{Spectrum of neutrino number flux $\Phi$ at Earth, distorted by $\nu$SI (red and blue line), assuming that without $\nu$SI, a blackbody flux is emitted (black line).}
    \label{fig:neutrino_spectrum}
%    \vskip-6pt
\end{figure}

The different spectra are shown in Fig.~\ref{fig:neutrino_spectrum}. Compared with the usual blackbody spectrum, $\nu$SI increase $\langle\epsilon\rangle$ by
10\% (19\%) if $\nu$SI do not conserve (do conserve) the neutrino number, and increase
$\langle\epsilon^2\rangle$ by 37\% (60\%). The impact of number-changing processes is very limited; particle number does not change in the fireball expansion, and only slightly changes at the emission from the neutrinosphere. In this sense, our results differ markedly from those of Ref.~\cite{Shalgar:2019rqe}, where only the $\nu\bar{\nu}\to\nu\bar{\nu}\nu\bar{\nu}$ reactions had been considered without accounting for the inverse reactions. In reality, the latter are as fast as the direct reactions, and at equilibrium they do not produce any large change in the average energy, lifting the bounds proposed in Ref.~\cite{Shalgar:2019rqe}. In the region of interest, for mediator masses above the MeV scale, only bounds based on laboratory, cosmology, and high-energy astrophysical neutrinos remain valid (see references in Ref.~\cite{Berryman:2022hds}). 

One observable we have not discussed is flavor. In the standard case, this issue is not fully understood, given the uncertainties on neutrino flavor conversion. On the other hand, assuming that all flavors are affected by $\nu$SI without overly hierarchical couplings, the spectra will be equalized, simplifying the situation somewhat.

{\bf\textit{Discussion and outlook.}}---We have investigated the emission, propagation, and observed spectrum of SN neutrinos, assuming they behave as a relativistic fluid caused by large $\nu$SI, drawing on theoretical foundations elaborated in our companion paper \cite{Fiorillo:2023cas}. We have used the simplest possible toy model of an isothermal and homogeneous source, but our results are generic up to factors of order unity. We find that the time profile is not modified by $\nu$SI---it is set by emission at the source, not by modified fluid propagation. On the other hand, the energy spectrum is somewhat harder than a blackbody spectrum, all else being equal. It is not obvious in which exact direction the neutrino energies and fluxes would change in a selfconsistent SN treatment. 

At present, we cannot estimate the realistic quantitative impact of different tens-of-percent effects due to $\nu$SI, as other modifications of this magnitude can play an important role~\cite{Bollig:2017lki, Melson:2015spa, Ehring:2023lcd, Ehring:2023abs}. Moreover, there is a region in the $g_\phi$--$m_\phi$ plane in which $\phi$ are trapped, but neutrinos do not behave as a fluid, and our results do not apply here. For the time being, while a future galactic SN might still be useful to unveil $\nu$SI, laboratory searches and future high-energy neutrino telescopes seem to be the best probe in large parts of the open parameter space~\cite{Berryman:2022hds}.

Treating neutrinos with large $\nu$SI as a relativistic fluid as pioneered by Dicus et al.\ \cite{Dicus:1988jh} has vastly simplified the discussion both conceptually and analytically. The \hbox{uncanny} smallness of the modifications caused by the fluid nature is the main surprise of this investigation and mandates a selfconsistent study to understand the exact quantitative effects in SN physics.

{\bf\textit{Acknowledgments.}}---We thank Shashank Shalgar, Irene Tamborra, Mauricio Bustamante, Po-Wen Chang, Ivan Esteban, John Beacom, Todd Thompson, Christopher Hirata, and Thomas Janka for informative discussions and/or comments on the manuscript.
DFGF is supported by the Villum Fonden under Project No.\ 29388 and the European Union's Horizon 2020 Research and Innovation Program under the Marie Sk{\l}odowska-Curie Grant Agreement No.\ 847523 ``INTERACTIONS.'' GGR acknowledges partial support by the German Research Foundation (DFG) through the Collaborative Research Centre ``Neutrinos and Dark Matter in Astro- and Particle Physics (NDM),'' Grant SFB-1258-283604770, and under Germany’s Excellence Strategy through the Cluster of Excellence ORIGINS EXC-2094-390783311. EV acknowledges support by the European Research Council (ERC) under the European Union’s Horizon Europe Research and Innovation Program (Grant No.\ 101040019). This article  is based upon work from COST Action COSMIC WISPers CA21106, supported by COST (European Cooperation in Science and Technology).

\bibliographystyle{bibi}
\bibliography{References}

\newpage

\onecolumngrid
\appendix

%\clearpage

\setcounter{equation}{0}
\setcounter{figure}{0}
\setcounter{table}{0}
\setcounter{page}{1}
\makeatletter
\renewcommand{\theequation}{S\arabic{equation}}
\renewcommand{\thefigure}{S\arabic{figure}}
\renewcommand{\thepage}{S\arabic{page}}

\renewcommand{\theequation}{A\arabic{equation}}
\renewcommand{\thefigure}{A\arabic{figure}}
\renewcommand{\thetable}{A\arabic{table}}
\setcounter{figure}{0} 
\setcounter{table}{0} 

%%%%%%%%%%%%%%%%%%%%%%%%%%%%%%%%%%%%%%%%%%%%%%%%%%%%%%%%%%%%%%%%%%%%%%%%%%%%%%%
%%%%%%%%%%%%%%%%%%%%%%%%%%%%%%%%%%%%%%%%%%%%%%%%%%%%%%%%%%%%%%%%%%%%%%%%%%%%%%%
\begin{center}
\textbf{\large Supplemental Material for the {\em Letter}\\[0.5ex]
Large Neutrino Secret Interactions, Small Impact on Supernovae}\\
\bigskip
Damiano G.~F.~Fiorillo, Georg G.~Raffelt, and Edoardo~Vitagliano
\end{center}
%%%%%%%%%%%%%%%%%%%%%%%%%%%%%%%%%%%%%%%%%%%%%%%%%%%%%%%%%%%%%%%%%%%%%%%%%%%%%%%
%%%%%%%%%%%%%%%%%%%%%%%%%%%%%%%%%%%%%%%%%%%%%%%%%%%%%%%%%%%%%%%%%%%%%%%%%%%%%%%

\begin{center}
\parbox{0.8\textwidth}{We summarize the hydrodynamical approach used in the main text. In addition, we consider the burst outflow scenario of Ref.~\cite{Chang:2022aas}, i.e., relativistic-hydrodynamical expansion for an initially homogeneous fluid ball at rest that is suddenly left to expand, a case for which these authors claimed homologous expansion. In complete contrast, we find instead that, even with these unrealistic initial conditions, the solution evolves toward a fireball (a~near-luminally expanding shell of constant thickness), in agreement with established wisdom and explicit earlier studies in a different context.}
\end{center}

\bigskip

\twocolumngrid

%\renewcommand{\tocname}{\vspace*{-0.8cm}}
%\tableofcontents

%%%%%%%%%%%%%%%%%%%%%%%%%%%%%%%%%%%%%%%%%%%%%%%%%%%%%%%%
%%%%%%%%%%%%%%%%%%%%%%%%%%%%%%%%%%%%%%%%%%%%%%%%%%%%%%%%

\section{1.~Evolution of a relativistic neutrino fluid in spherical symmetry}

Due to the assumed short neutrino-neutrino mean free path, a kinetic description of neutrino emission from supernova is inadequate. A more appropriate fluid description must be constructed from the hydrodynamical equations of a fluid with a relativistic equation of state, i.e., $p=\rho/3$, where $p$ is the rest-frame pressure and $\rho$ the rest-frame energy density. For generality, let us assume that the fluid can exchange energy and momentum with a surrounding medium such as the dense matter inside a protoneutron star (PNS). The general hydrodynamical equations of motion then are 
\cite{Weinberg:1972kfs}
\begin{equation}\label{eq:fluid_eqs}
    \partial_\nu T^{\mu \nu}=S^\mu, 
\end{equation}
where
\begin{equation}
    T^{\mu\nu}=(\rho+p)\,u^\mu u^\nu-p\,g^{\mu\nu}=\frac{4}{3}\rho\,u^\mu u^\nu-\frac{\rho}{3}g^{\mu\nu}
\end{equation}
is the stress-energy tensor for a fluid with bulk four-velocity $u^\mu$, whereas $S^\mu$ is the energy-momentum exchanged with the medium per unit time and volume. If this term vanishes, as for free expansion in space, these equations simply express the local conservation of energy and momentum. 

If neutrino-neutrino interactions also cause fast num\-ber-changing reactions such as $\nu\bar{\nu}\to\nu\bar{\nu}\nu\bar{\nu}$, the neutrino gas in the comoving fluid frame will relax to an isotropic distribution in thermal and chemical equilibrium. In this limit, the rate of direct and inverse reactions are always equal. Therefore, the bounds proposed in Ref.~\cite{Shalgar:2019rqe}, which considers only the direct reactions but neglect the inverse ones, cannot apply.

If instead number-changing reactions are too slow, the number of neutrinos and antineutrinos is conserved by the interaction, and therefore the hydrodynamical equations must be complemented by an equation for particle number conservation
\begin{equation}
    \partial_\mu (n u^\mu)=Q,
\end{equation}
where the term on the right-hand side describes the number of neutrinos exchanged per unit volume and unit time with the medium. We emphasize that, whether number conservation holds or not, the equations for energy and momentum conservation Eq.~\eqref{eq:fluid_eqs}, and therefore the energy density and bulk velocity, are identical. 

For spherically symmetric motion, these equations considerably simplify. The bulk velocity can now be expressed in terms of a single radial velocity $v$, such that $u^0=\gamma=(1-v^2)^{-1/2}$ and $u^r=\gamma v$. The lab-frame energy density $e$ is
\begin{equation}\label{eq:lab-energy}
    e=\frac{4\gamma^2-1}{3}\,\rho=\frac{1+v^2/3}{1-v^2}\,\rho,
\end{equation}
and the lab-frame momentum density $T^{0r}$ is
\begin{equation}
    T^{0r}=e\xi=\frac{4}{3}\rho\gamma^2 v.
\end{equation}
Here we have introduced the modified velocity variable 
\begin{equation}
    \xi=\frac{4\gamma^2v}{4\gamma^2-1}=\frac{4 v}{3+v^2}
\end{equation}
extensively used in our companion paper~\cite{Fiorillo:2023cas}. This new velocity was defined precisely such that the fluid (energy flux) = (energy density $e$) $\times$ (modified velocity $\xi$). 

The specific form of the energy-momentum exchange term $S^\mu$ depends on the type of interaction with the medium. For the case of an energy-independent mean free path $\lambda_{\nu N}$ for absorption by the medium, we have shown in our companion paper that the transfer rates of energy, momentum, and number density are~\cite{Fiorillo:2023cas}
\begin{equation}
    S^0=\frac{e_{\rm eq}-e}{\lambda_{\nu N}},
    \quad
    S^r=-\frac{e\xi}{\lambda_{\nu N}},
    \quad
    Q=\frac{N_\mathrm{eq}-N}{\lambda_{\nu N}}, 
\end{equation}
where $e_{\rm eq}$ is the fluid energy density in the lab frame if it were in local thermal equilibrium with the medium, $N$ the fluid particle number density in the lab frame, and
$N_{\rm eq}$ its value in local chemical and thermal equilibrium with the medium. 

\begin{widetext}
With this result, we can readily write down the full system of hydrodynamical equations
\begin{subequations}\label{eq:sphere_time_equation}
\begin{eqnarray}\label{eq:sphere_time_equation-a}
&&\partial_t e\kern1.2em
+\frac{\partial_r(e\xi r^2)}{r^2}\kern18.9em
=\frac{e_{\rm eq}-e}{\lambda_{\nu N}},
\\[1.5ex]
\label{eq:sphere_time_equation-b}
&&\partial_t(e\xi)+\frac{1}{3r^2}\partial_r\left[e\left(5-2\sqrt{4-3\xi^2}\right) r^2\right]
-\frac{2e}{3r}\left(\sqrt{4-3\xi^2}-1\right)
=-\frac{e\xi}{\lambda_{\nu N}},
\\[1.5ex]
\label{eq:sphere_time_equation-c}
&&\partial_t N\kern0.8em
+\frac{\partial_r(N v r^2)}{r^2}\kern18.4em
=\frac{N_\mathrm{eq}-N}{\lambda_{\nu N}},
\end{eqnarray}    
\end{subequations}
where $\sqrt{4-3\xi^2}=2\,(3-v^2)/(3+v^2)$. This set of equations must be complemented with initial conditions to obtain the solution to a physical problem. In our main text, we consider the solution for a spherical blackbody with temperature~$T$, simulating neutrino injection from the central PNS.
\end{widetext}

These equations correspond to Eqs.~(H11)--(H13) of Ref.~\cite{Chang:2022aas} that were written in terms of the comoving fluid density which they call $\tilde\rho$, but we call $\rho$. In other words, inserting Eq.~\eqref{eq:lab-energy} in our equations and in addition set the right-hand side to zero for the evolution in free space recovers those in Ref.~\cite{Chang:2022aas}. We are certainly solving the same equations.

\section{2.~Free expansion of a neutrino fluid}

The main idea of Ref.~\cite{Chang:2022aas} to obtain large effects on a SN neutrino signal was the hypothesis that the fluid nature causes the neutrino burst to stretch in time. While they do consider an alternative case, namely the so-called wind outflow, which is a stationary solution of the fluid equations, they do not provide any reason either in favor or against its relevance. In fact, solving the steady equations cannot by itself provide information on any observable. Our dynamical treatment shows that the emission relaxes to a steady behavior only during the PNS cooling phase, and only close to the PNS surface. The behavior at the beginning and at the end of the cooling phase, and far from the PNS, can only be addressed by a dynamical treatment the we have performed.

As an alternative possibility to the wind outflow, they propose the burst outflow solution based on the scenario of an initial spherical ball of neutrino fluid suddenly left to expand. We have dismissed this solution in the main text because such an initial condition looks unphysical and Ref.~\cite{Chang:2022aas} did not explain how it might be produced by SN-related physics. One may still ask the abstract question of how such an initial condition would evolve, and indeed this question has been addressed a long time ago in a classic paper~\cite{vitello1976hydrodynamic} to which we will return later. In any event, there should be a simple answer based on Eqs.~\eqref{eq:sphere_time_equation-a} and \eqref{eq:sphere_time_equation-b} with vanishing right-hand side and solving for the unknown functions $e(r,t)$ and $\xi(r,t)$.

As a first step one may search for a simple analytic solution, where a homologous ansatz could be promising, meaning solutions that depend as power laws on $r$ and $t$ and are thus scale free. Different families of such solutions are conceivable and we here follow Ref.~\cite{Chang:2022aas} and assume as the simplest case
\begin{equation}\label{eq:velocity}
    v(r,t)=r/t
    \quad\hbox{for}\quad r\leq t.
\end{equation}
The point $r=t$ corresponds to a luminally expanding shock front; there is no fluid beyond this point. The corresponding lab-frame energy density is
\begin{equation}
    e(r,t)\propto \frac{3t^2+r^2}{(t^2-r^2)^3}.
\end{equation}
One can explicitly check that these functions indeed solve
Eqs.~\eqref{eq:sphere_time_equation-a} and \eqref{eq:sphere_time_equation-b} with vanishing right-hand side.

These solutions do not depend on the particle number density. With the linear velocity profile of Eq.~\eqref{eq:velocity}, the third equation Eq.~\eqref{eq:sphere_time_equation-c} with vanishing source on the right-hand side is solved by
\begin{equation}
    N(r,t)\propto \frac{1}{t^3}\phi\left(\frac{r}{t}\right)
\end{equation}
for any function $\phi(x)$. In particular, if the lab-frame number density was initially set up to be homogeneous, it will stay that way, but decreasing (diluting) with $t^{-3}$.

While this solution may at first seem suggestive, it is actually pathological in that the energy density is extremely singular at the edge of the expanding fluid at $r=t$. The total contained energy is infinite and concentrated in an infinitely thin layer at the surface. This issue was already raised a long time ago by Ref.~\cite{Blandford:1976uq}, who first obtained this solution. If one were to assume a regularized edge of the fluid, this would break the scale invariance, and it is no longer clear if such a modified profile would solve the equations of motion. Moreover, the homologous solution does not connect with an assumed initial condition of a fluid at rest that is suddenly left to expand.

Finally, because the assumed number density and energy density are separate solutions, the assumed energy per particle becomes singular toward the edge of the fluid if one were to assume a homogeneous number density and this would have to be assumed as an initial condition. At a given comoving radius, for example $r=t/2$, the lab-frame energy density decreases as $t^{-4}$, the particle number density as $t^{-3}$, and so the energy per particle as $t^{-1}$, similar to the expanding universe.

\begin{figure*}
    \includegraphics[width=0.42\textwidth]{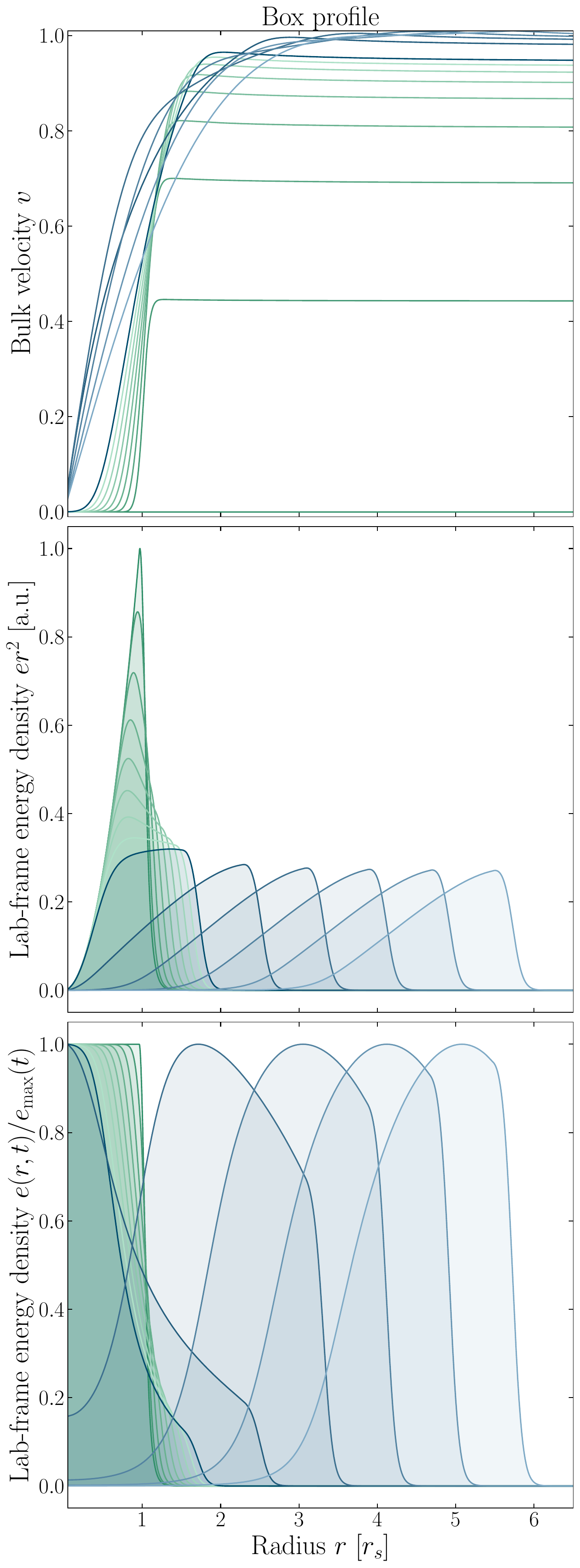}
    \includegraphics[width=0.42\textwidth]{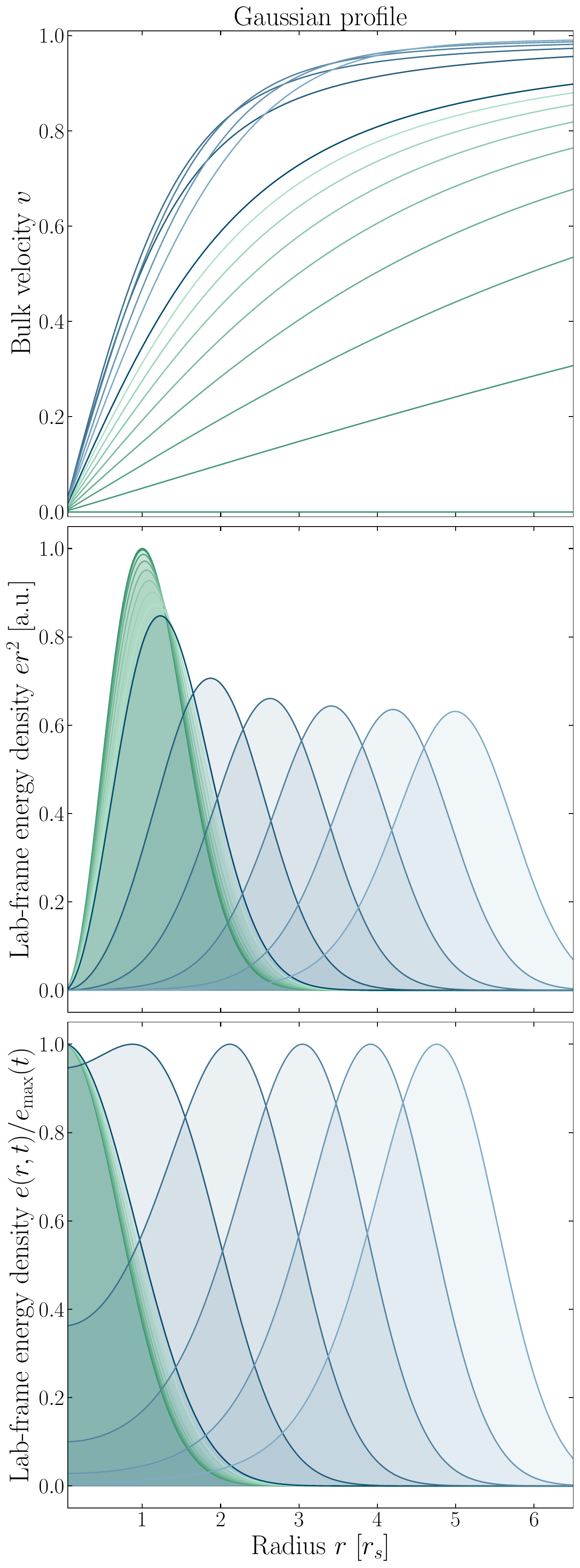}
    \caption{Lab-frame energy density and velocity profile for the evolution of a ball of neutrino fluid at rest, for the box (left) and gaussian (right) profile. We show velocity (top), lab-frame energy density times $r^2$ (middle), and lab-frame energy density normalized to its maximum value at each time (bottom). The progressively lighter shadings in green correspond to snapshots in time taken every $0.1\;r_s$ from $0\;r_s$ to $0.7\;r_s$; the blue shadings correspond to snapshots in time taken every $0.8\;r_s$ from $0.8\;r_s$ to $4.8\;r_s$.
    }   
    \label{fig:free_expansion}
\end{figure*}

To understand if an initial fluid ball at rest would evolve toward something reminiscent of this interesting solution, one can directly solve Eqs.~\eqref{eq:sphere_time_equation-a} and \eqref{eq:sphere_time_equation-b} numerically using a simple finite difference method.\footnote{We have also succeeded to solve these partial differential equations with the standard routine {\tt NDSolve} in {\sc Mathematica}, taking the radial range 0--$8\,r_{\rm s}$ and $t$ up to around $6\,r_{\rm s}$, more than enough to see the initial fluid ball to morph to a fireball. Therefore, one does not need specialized codes to gain a clear numerical picture.}  We consider as initial condition a ball of neutrino fluid at rest. The initial energy density profile is chosen to be either homogeneous within a radius $r_s$ (box profile), or Gaussian with $\rho(0,r)\propto \exp\left[-r/r_s\right]$. For the box profile, we consider an exponential drop to zero after $r>r_s$ as $\exp\left[-20\;r_s^{-1}\;(r-r_s)\right]$. By testing both profiles, we make sure that our results are not determined by sharp edge of the box profile. Our numerical solutions are shown in Fig.~\ref{fig:free_expansion}.

For both density profiles, the interface with the vacuum causes the neutrino fluid at the surface to move outward. In the box profile, we can even see how initially the fluid deep inside the ball remains unaffected, since the information about the interface with the vacuum takes a finite time $\sim r_s$ to reach the inner region of the ball. Therefore, in this case, a rarefaction wave is launched from the surface towards the interior of the ball, causing the fluid to progressively start moving outwards over a typical time $r_s$.

After the fluid has left the initial ball, it settles into a shell with a velocity profile that is not linear. Rather, the velocity profile grows much more rapidly, and, in both cases, we find that the bulk of the neutrino fluid shell lies in a region where the velocity is close to the speed of light. Therefore, the thickness of the fluid shell does not further increase. In other words, the solution for the energy density evolves toward a fireball solution, in particular with the property of having a scale imprinted (the original radius of the ball). Already from a time of about $2.4$~$r_s$, the quantity $e r^2$ maintains an identical profile and simply moves uniformly with the speed of light, as predicted from the fireball solution presented in the main text. Therefore, we find that, independently of the precise form of the initial density profile, a ball of neutrino fluid initially at rest quickly evolves into a fireball with constant thickness.

These initial conditions have been studied in quite a different context, namely that of gamma-ray bursts (GRBs). We can trace the first analysis of a similar setup to Ref.~\cite{vitello1976hydrodynamic}, which solved by a different numerical method the identical problem of the box profile for a generic relativistic fluid. The results found there agree with ours, and in fact Ref.~\cite{vitello1976hydrodynamic} concludes that after an initial readjustment, clearly visible also in our Fig.~\ref{fig:free_expansion}, there is no change in the thickness of the plasma shell. These simulations have played an important role in the fireball model of GRBs, in which an energy injection in the form of a relativistic fluid in a small region is usually assumed, see, e.g., Ref.~\cite{Piran:1993jm}. 

The setup of the GRB fireball is not exactly identical to our problem, since in that case the relativistic fluid has a small contamination of non-relativistic baryons and it expands into an external non-relativistic medium rather than into the vacuum. However, the very early stages of expansion, when the energy of the fireball is radiation-dominated, are analogous to the problem studied here. Ref.~\cite{Piran:1993jm} provides numerical simulations for a regularized box profile [see their Eq.~(14)], and also concludes, in agreement with Ref.~\cite{vitello1976hydrodynamic} and with our own results, that in the radiation-dominated phase the shape of the pulse is not changed.

By itself, the fact that the thickness of the neutrino shell remains unaltered by the expansion may be a source of confusion, especially if we are used to thinking of non-relativistic fluids and non-relativistic bulk velocities. For the latter, it is indeed true that an expansion always obtains. If we set up a shell of fluid moving radially with non-relativistic velocity, the lack of pressure support both at the front and at the back of the fluid shell causes it to expand at both sides and ultimately evolve into a homologous expansion (see, e.g., Ref.~\cite{zeldovich1966physics}).
Why is this not true for relativistic fluids? A qualitative explanation is provided in Ref.~\cite{vitello1976hydrodynamic}. In the rest frame of the shell of fluid, it is indeed true that the shell expands, driven by the lack of pressure support. However, passing to the lab-frame requires a strong boosting, so that the expansion is slowed down by a factor $1/\gamma^2$. Since the fluid shell soon is entrapped in a motion with $\gamma\propto r$, it is clear that 
the growth of thickness
is rapidly slowed to a halt.

Our findings are in complete contrast to what is claimed in Ref.~\cite{Chang:2022aas}. Concerning their homologous analytic solution, it formally solves the equations of motion, but is extremely singular and requires being set up in exactly this form, a question not addressed in Ref.~\cite{Chang:2022aas}. These authors also claim that an initial setup of the relativistic fluid at rest would generically evolve toward something similar to the homologous solution, but with a smoothed edge. Details of their numerical solutions with the PLUTO code are not documented, so one cannot judge what exactly was done or found. For comparison, they cite Ref.~\cite{yokosawa1980relativistic}, which however performs numerical simulations for the plane-parallel problem only. In fact, a discussion of the spherical expansion presented in Ref.~\cite{yokosawa1980relativistic}, based on analytical arguments, fully supports the conclusions of Ref.~\cite{vitello1976hydrodynamic}, and therefore contradicts rather than supports the findings of Ref.~\cite{Chang:2022aas}.

Their claim that an initial fluid ball at rest would expand in a homologous fashion therefore contradicts established wisdom and our finding of an evolution toward a fireball. Thus, we conclude that, even for the initial conditions assumed in Ref.~\cite{Chang:2022aas}, a change in the duration of the neutrino signal is not expected, and the burst outflow does not obtain. On the contrary, both types of initial conditions, and with the more realistic ones used in the main text and corresponding to injection from the PNS surface, the solutions always evolve toward a fireball with a frozen pulse profile, allowing us to robustly conclude that this feature does not strongly depend on the mechanism by which neutrinos are injected.

\end{document}